# Rapid Application Development Using Software Factories

*Toni Stojanovski PhD, Tomislav Dzekov PhD*
Faculty of Informatics, European University, Republic of Macedonia
toni.stojanovski@eurm.edu.mk, tomislav.dzekov@eurm.edu.mk

***Abstract:*** *Software development is still based on manufactory production, and most of the programming code is still hand-crafted. Software development is very far away from the ultimate goal of industrialization in software production, something which has been achieved long time ago in the other industries. The lack of software industrialization creates an inability to cope with fast and frequent changes in user requirements, and causes cost and time inefficiencies during their implementation.*

*Analogous to what other industries had done long time ago, industrialization of software development has been proposed using the concept of software factories. We have accepted this vision about software factories, and developed our own software factory which produces three-layered ASP.NET web applications.*

*In this paper we report about our experience with using this approach in the process of software development, and present comparative results on performances and deliverables in both traditional development and development using software factories.*

## 1. INTRODUCTION

Today's enterprise software solutions operate in a very volatile environment. Often, as a result of change requests and additional requirements that were not known at the beginning of the project, the project scope grows by 20-100%. In order to graciously accept these changes in requirements, the focus of the software development practices during the last three decades has been constantly shifting towards the design of a system that can withstand unpredictability. From a user's perspective, software is required to be not only reliable, but also up-to-date with the technical and functional requirements. Software project's final delivery does not mark the closure of the solution's lifecycle. Maintenance and upgrading still account for two thirds of software's lifetime cost. For the above mentioned reasons, software needs to be adaptable in a cost-effective manner to the following three types of changes:

- **Functional changes** add new capabilities to the software while retaining the existing platform and performance.
- **Non-functional changes** address performance issues e.g. availability, reliability, response times, security etc. without changes to the platform base.
- **Platform changes** are a result of moving the software to new hardware or

operating system, or using new or modified services.

Despite of the increase in reuse and productivity of software development and the quality of software products brought by the object-oriented programming and service-oriented architecture during the past two decades, there is still demand for further improvements. Software engineering practices are still in the craftsmanship age. As a result, software development is slow and expensive; yields products containing serious defects that cause problems of usability, reliability, performance, and security; and has excessive reliance on the human factor.

Furthermore, most software is still built more or less in isolation. Reuse is mostly limited. However, if we analyse the applications we can often identify a large amount of similar functionalities that is developed from scratch for each application. Developing applications often involves repetitive coding. Developers are rewriting code somebody else or even themselves have already written before, and are duplicating the same segments of code throughout the software solution. Reducing repetitive coding has been the focus of software engineering since its birth.

The major reason for the above issues is the fact that the software development has not been industrialized yet. Most software today is still being developed by hand from scratch using labour-intensive methods. Although in the software industry there have been individual attempts to accept practices of other industries over the past 30 years, the analysis show that there are only a small number of cases where software development is automated. Furthermore, there is no single widely accepted software factory-based methodology. Software engineering is now at a breaking point where it must adopt the patterns of industrialization demonstrated by other industries. These include assembling products from components, automating repetitive or menial tasks, forming product lines, and standardizing processes, architectures, and packaging formats.

According to [1], "software factory (SF) is a configuration of languages, patterns, frameworks, and tools that can be used to rapidly and cost-effectively produce an open-ended set of unique variants of a standard product." Software factories are also sometimes called Generative Programming [2] and Code Generation [3, 4]. The SF approach adopts the patterns of industrialization demonstrated by other industries. In layman terms, an SF can produce software solutions similar to the way a car factory produces cars, and a jam factory produces jam. Clearly, car factories can produce other types of vehicles, that is, products from the same family, but no jam. To continue with the above analogy, an SF can create software products from one family only e.g. web portals, but another SF is required to produce solutions from another family e.g. document management software.

Motivated by the above mentioned problems in software engineering, we have built a software factory which automates the development of three-tiered ASP.NET web applications. This paper describes the implemented software factories, and the impact it has on the software design and developers' productivity.

We also note that there are other approaches to the problems of changes in software requirements, repetitive coding, and insufficient reuse, such as Agile methodology [5] (Scrum [6], Extreme Programming [7] etc.), Rapid prototyping [8], Model-driven architecture [9], and Design patterns [10]. These approaches are not mutually exclusive with the use of SFs, and can be used symbiotically.

Here is the overview of our paper. Section 2 describes the concept and main components of SF. Section 3 describes the phases in the development of our SF. Section 4 describes the benefits and advantages that SF brings to software development. Section 5 presents code statistics. Section 6 gives the conclusion, and the directions for our future work.

## 2. IMPLEMENTED SOFTWARE FACTORY

Software factories are about modelling and implementing system families in such a way that a given system can be automatically generated from a specification written in a domain-specific language (DSL). SF means using a program to help you write your programs. SF reads a specification of abstract requirements , and then using templates builds one or more output files for the software solution.

Our software factory consists of three major components, and many other SFs consist of the same components.

- A DSL that can be used to define a formal model/blueprint of the application. The DSL is used to define the software from a business point of view. Ideally, the same DSL-written application model can be processed by two sets of artefact templates to generate two solutions for different platforms e.g. .NET and J2EE. Our DSL is based on XML.
- An approved software architecture which satisfies the technical requirements for the application. The software architecture is defined by means of a set of artefact templates written in XSLT. These templates transform the application model into a number of artefacts: SQL scripts, documentation, help files, source code etc.
- A software tool (code generator) that can apply the artefact templates on the application model, and produce the entire software application. Our software tool is named RoboCod, and works as an add-on for MS Visual Studio 2005 and later versions.

The following figure gives the software factory workflow:

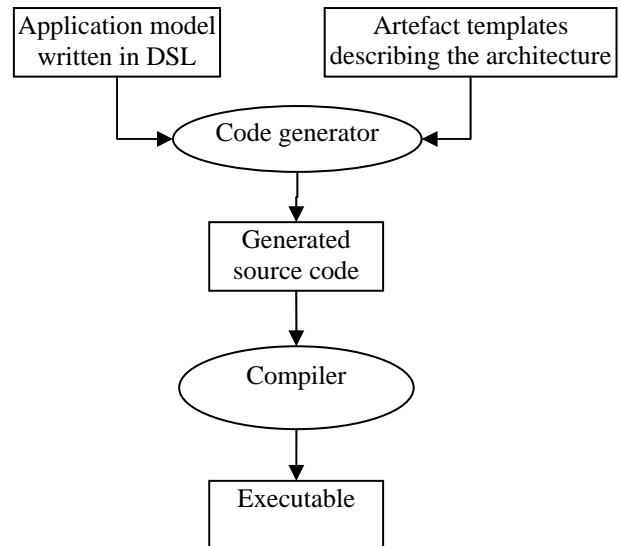

*Figure 1. Software factory workflow*

We use the best practices of software reuse and design patterns during the SF development. Our solution is based on standardised languages: Extensible Markup Language (XML) and Extensible Stylesheet Language Transform (XSLT) [13].

An XSLT is a program to transform XML input document into text-based output using optional input parameters. The benefits of using XSLT are as follows:

- Can generate anything text-based;
- Standards-based;
- Visual Studio 2005 and later versions provide an XSLT debugger.

In our opinion the only drawback of using XSLT is that it uses declarative programming, which can be difficult to learn.

When talking about the scope and domain of SF, we can distinguish between two types of domains in which the factory can belong:

- **Vertical** - targets software products that have a precisely defined business domain. The applications which will be produced with SF's from such a domain will have a common business focus, i.e. applications for insurance policies.
- **Horizontal** –targets software products that have a common architecture e.g. three-layered distributed application, or operating platform e.g. J2EE, .NET

In other words, Domain Model does vertical focusing of the SF, while Artefact Templates do horizontal focusing.

Our SF is a combination of the two domains: it produces three-layered ASP.NET web application with a set of supported business rules and constraints. We can reuse the SF to generate another application from another business domain provided that its business rules can be described in the terms of the SF's existing business rules, and it has the same architecture. As a consequence of this flexibility, our SF has been used to implement a fee calculation application for financial institutions, University News-Board web application for university, phone-list application, and issue tracker for software development organisations.

## 3. BUILDING A SOFTWARE FACTORY

Before the application development takes place, the SF must be built. Clearly, once an SF is built, it can be used again and again to build many applications. Our SF was built in four phases.

The first phase is development of the domain vocabulary and grammar of the dictionary that covers the business terms of the domain area where the application that will be produced by the SF belongs. For this purpose, we have developed a DSL based on XML. Using XML allows us to reuse its grammar implementation, and to avoid the implementation for a new grammar, Furthermore, we benefit from its standardisation, abundance of editors and debuggers, wide acceptance, and its familiarity to programmers. We use XML schema [11, 12] definition to validate application models, which is another benefit of using XML. Our DSL is platform agnostic. The same DSL-written application model can be processed by many sets of artefact templates to produce two solutions for different platforms e.g. .NET and J2EE. Although the initial Domain Model can be reused in further application development without any modifications, often there is a need to change or extend the Domain Model in order to accommodate the new requirements from the already covered business domains, or the introduction of new business domains. Following figure gives a fraction of the XML schema, which defines the definition for a field in an entity.

```
<xs:element maxOccurs="unbounded" name="Field" minOccurs="1">
    <xs:complexType>
    <xs:attribute name="name" type="xs:string" use="required" />
    <xs:attribute name="isShownInList" type="xs:boolean" use="optional" />
    <xs:attribute name="isShownInEdit" type="xs:boolean" use="optional" />
    <xs:attribute name="isIdentity" type="xs:boolean" use="optional" />
    <xs:attribute name="isPK" type="xs:boolean" use="optional" />
    <xs:attribute name="type" type="FieldType" use="required" />
    <xs:attribute name="description" type="xs:string" use="optional" />
    <xs:attribute name="nullable" type="xs:boolean" use="optional" />
    <xs:attribute name="displayName" type="xs:string" use="optional" />
    <xs:attribute name="defaultValue" type="xs:string" use="optional" />
    <xs:attribute name="displayFormat" type="xs:string" use="optional" />
    <xs:attribute name="length" type="xs:string" use="optional" />
    <xs:attribute name="isLookup" type="xs:boolean" use="optional" />
    <xs:attribute name="isOVN" type="xs:boolean" use="optional" />
    <xs:attribute name="isFK" type="xs:boolean" use="optional" />
    <xs:attribute name="nameName" type="xs:string" use="optional" />
    <xs:attribute name="fkEntityName" type="xs:string" use="optional" />
    <xs:attribute name="createLookup" type="xs:boolean" use="optional" />
    <xs:attribute name="isAudited" type="xs:boolean" use="optional" />
    <xs:attribute name="numberOfRows" type="xs:unsignedByte" use="optional" />
```

```
        <xs:attribute name="numberOfCols" type="xs:unsignedByte" use="optional" />
        <xs:attribute name="isShownInHistory" type="xs:boolean" use="optional" />
        </xs:complexType>
</xs:element>
```

*Figure 2. A fraction of the XML schema for the DSL.*

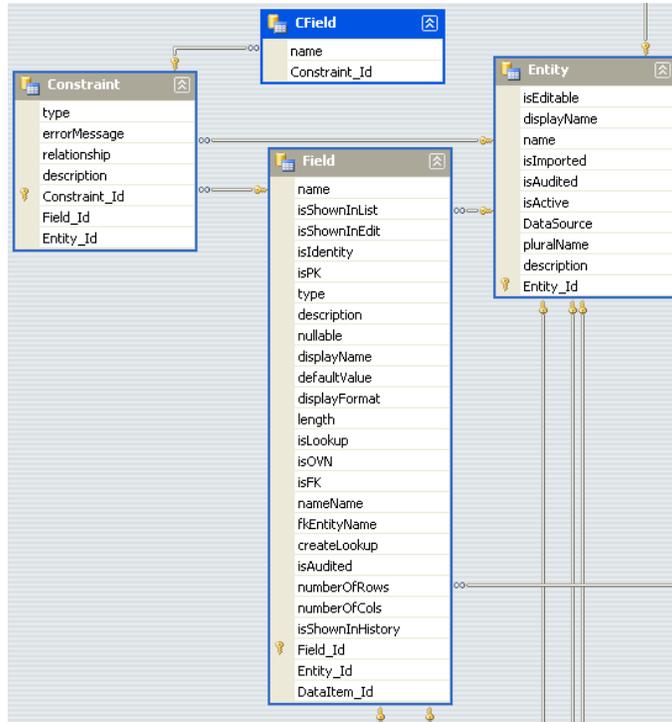

*Figure 3. A fraction of a graphical presentation of the XML schema for the DSL.*

In the next phase we define a DSL-based Domain Model which contains application's details, such as entity-relationship model, business constraints, connection strings, as well as user interface functionality and navigation.

Following figure gives a fraction of the DSL-based domain model for the University News-Board web application.

```
<Entity tableName="Fakultet" name="Fakultet" caching="enabled" isAudited="true" isLogged="true"
isActive="true">
    <Language name="Macedonian">
        <DisplayName>Факултет</DisplayName>
        <PluralName>Факултети</PluralName>
    </Language>
    <Language name="English">
        <DisplayName>Faculty</DisplayName>
        <PluralName>Faculties</PluralName>
    </Language>
    <Field name="ID" isShownInList="false" isIdentity="true" isPK="true" type="int"
description="Record ID" nullable="false">
```

```
        <Language name="Macedonian">
            <DisplayName>ИД</DisplayName>
        </Language>
        <Language name="English">
            <DisplayName>ID</DisplayName>
        </Language>
    </Field>
    <Field name="strName" type="nvarchar" length="30" nullable="false">
        <Language name="Macedonian">
            <DisplayName>Име</DisplayName>
        </Language>
        <Language name="English">
            <DisplayName>Name</DisplayName>
        </Language>
    </Field>
...
    <Constraint type="Unique">
        <Language name="Macedonian">
            <ErrorMessage>Името на Факултетот мора да биде уникатно.</ErrorMessage>
        </Language>
        <Language name="English">
            <ErrorMessage>Faculty name must be unique</ErrorMessage>
        </Language>
        <CField name="strName" />
    </Constraint>
</Entity>
```

*Figure 4. A fraction of the DSL-based domain model for the University News-Board web application.*

The third phase is the most difficult and complex part of building web-based application. The application architecture needs to be prepared. After that, the code for one entity is written and then it is generalised through templates. The act of templating the existing, working code for reuse is also called code generalization.

```
<xsl:template match="/">
  <xsl:for-each select="xsource/EntityConfig/Entity[@isActive='true']">
    CREATE TABLE [dbo].[tbl_<xsl:value-of select="@tableName"/>] (
    <xsl:call-template name="TableColumns"/><xsl:if test="@isLogged='true'">,    [changedAt]
[datetime] NOT NULL,
    [changedBy] [varchar](50) NOT NULL</xsl:if>
    ) ON [PRIMARY]
    GO
  </xsl:for-each>
</xsl:template>
<xsl:template name="TableColumns">
  <xsl:for-each select="Field">
  [<xsl:value-of select="@name"/>] <xsl:call-template name="ModelTypeToDB"/><xsl:call-template
name="Collation"/><xsl:choose><xsl:when test="@nullable='true'">
NULL</xsl:when><xsl:otherwise> NOT NULL</xsl:otherwise></xsl:choose><xsl:if
test="@isIdentity='true'"> IDENTITY (1, 1)</xsl:if><xsl:if test="position()!=last()">,</xsl:if>
```

```
</xsl:for-each>
</xsl:template>
```

*Figure 5. A fraction of the XSLT template which generates a CREATE TABLE SQL script.*

We have written artefact templates that generate the code for the following tiers:

- Data Base: These templates produce SQL scripts to build database tables, constraints, views, indexes, stored procedures for MS SQL Server 2000 and MS SQL Server 2005.
- Data Access Components (DAC): They abstract the logic necessary to access data in a separate set of data access components. They will provide Create, Read, Update and Delete operations to specific business components. The Data Access Components will help to reduce code duplication, provide for data access tuning and optimisation as well as application manageability.
- Business Components: These templates produce Business Components that encapsulate the business logic and rules for each of the business entities.
- Business Layer Service Interface (BLSI): These templates will produce Business Layer Service Interface that will expose the business logic to the presentation layer. The BLSI shields the business logic and allows for changes in business logic in future making the code maintainable, and decouples the business logic from presentation layer. The BLSI is the façade for the business logic.
- Web User Interface: These templates are the most complex compared to the previous ones. These templates produce the presentation layer which consists of a set of ASP.NET Web Forms. These Web Forms will be responsible for displaying the application data to the users, and for collecting the user input.

Furthermore, we have also developed templates that generate

- Data Transfer objects as generic and strongly typed datasets,
- Web Services,
- Reports in report definition language –client (.rdlc),
- Documentation and online help.

In the last phase, we developed a Software Generator Tool as an add-on for Visual Studio 2005 and Visual Studio 2008, named RoboCod. RoboCod uses the Domain Model written in DSL and the Artefacts Templates corresponding to the technical architecture as two inputs. It produces the code as a one-click activity. The code is ready to be compiled. The produced code is ASP.NET code which together with the handcrafted code is compiled.

Even in the most typical representative of a family of software applications there will be atypical bits of functionality, which cannot be described in the DSL. This functionality still needs to be handwritten by developers. As a rule of thumb, if a business rule appears in less than three entities, then it is more efficient to handcraft the code rather than to model business rule in the DSL and then modify the templates for the artefacts. For less than three entities, building the templates is not recommended. With respect to the handcrafted code, it is very important to preserve the hand written changes to the SF generated code.

Figure 6 shows the class hierarchy which allows the shared ownership and protection of handcrafted code.

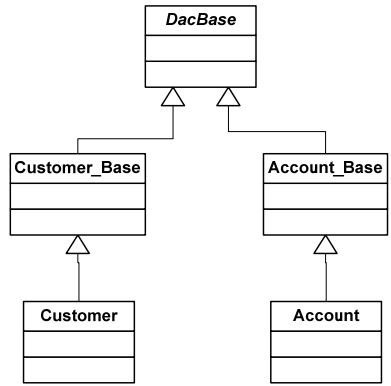

*Figure 6. Class hierarchy which allows shared ownership.*

- *DacBase* is an abstract class that defines the base interface and data members for all the DACs.
- *Customer_Base* class implements the methods defined in *DacBase*. This class and every other *<EntityName>_Base* class is created and owned by SF, that is, any changes done manually to this class can be overwritten by SF.
- *Customer* class inherits from *Customer_Base* and contains no implementation. It is created (only once) by SF, but it is not owned by SF, that is, any changes done manually to this class will not be overwritten by SF. This class can be used to override the default functionality implemented in *Customer_Base*, if such a need exists.

C# 2.0 feature of partial classes is particularly helpful when we need to have shared ownership over ASP.NET forms, because part of a class can be SF produced while leaving other parts of the class to be written by hand.

## 4. DESIGN IMPACTS

SFs bring following benefits and advantages to software development:

- **Quality:** SFs use templates to build the code. Improving the quality of the templates is directly translated into increased quality of the overall code base.
- **Horizontal Consistency:** Since the application model is the central and only repository of the functional specification, all the artefacts (code from multiple tiers, documentation, test scripts, help files, deployment scripts etc.) which are automatically produced from the application model will be consistent with each other and up-to-date with the latest version of the model. When SFs are not used, then code improvements which are discovered lately in the project are too expensive and risky to be implemented throughout the whole solution. Consequently, early written code will not contain such code improvements and there will be inconsistent implementations for the same business rules and problems. However, when SFs are used, code improvements discovered lately in the project can still be implemented in the artefact templates, and then can be replicated to the entire solution.

- **Consistent implementation of business rules (Vertical consistency)**: When SFs are not used, the tendency is to have the business rules implemented in only one tier (typically in the business components) in order to avoid the inconsistency of implementations of business rules across different tiers. However, with SFs, the definition of business rules is centralised in the **application model,** not in the **business components**. Then, SF spreads the implementation of the business rules to multiple tiers. Thus, the validation can be done in multiple tiers (validation-in-depth) and the consistency of the business rules is enforced by the SF.

```
<Constraint type="TwoFields" relationship="le">
    <CField name="DisplayFrom" />
    <CField name="DisplayTo" />
</Constraint>
```

*a) Definition of a business rule.*

```
<xsl:for-each select="Constraint[@type='TwoFields']">
if (validation.vs_compare_<xsl:choose><xsl:when
test="../Field[@name=current()/CField[position()=1]/@name][@type='datetime' or
@type='date']">dates</xsl:when><xsl:otherwise>strings</xsl:otherwise></xsl:choose><xsl:call-
template
name="IsRelationshipNullable"/>(aspnetForm.ctl00_MainContentplaceholder_ctrl<xsl:value-of
select="CField[position()=1]/@name"/>, aspnetForm.ctl00_MainContentplaceholder_ctrl<xsl:value-of
select="CField[position()=2]/@name"/>, '<xsl:value-of select="@relationship"/>'))</xsl:for-each>
```

*b) XSLT template which implements the business rule in the web presentation tier.*

```
<xsl:for-each select="Constraint[@type='TwoFields']">
ALTER TABLE [dbo].[tbl_<xsl:value-of select="$TableName"/>] ADD
CONSTRAINT [CK_tbl_<xsl:value-of select="$TableName"/>_<xsl:value-of
select="CField[position()=1]/@name"/>_<xsl:value-of select="CField[position()=2]/@name"/>]
CHECK ([<xsl:value-of select="CField[position()=1]/@name"/>] <xsl:choose>
    <xsl:when test="@relationship='lt'"><</xsl:when>
    <xsl:when test="@relationship='le'"><=</xsl:when>
    <xsl:when test="@relationship='gt'">></xsl:when>
    <xsl:when test="@relationship='ge'">>=</xsl:when>
    <xsl:when test="@relationship='neq'"><></xsl:when>
    <xsl:when test="@relationship='eq'">>=</xsl:when>
</xsl:choose> [<xsl:value-of select="CField[position()=2]/@name"/>])
GO
</xsl:for-each>
```

*c) XSLT template which implements the business rule in the data tier.*

```
if (validation.vs_compare_dates(aspnetForm.ctl00_MainContentplaceholder_ctrlDisplayFrom,
aspnetForm.ctl00_MainContentplaceholder_ctrlDisplayTo, 'le'))
```

```
ALTER TABLE [dbo].[tbl_Vest] ADD
CONSTRAINT [CK_tbl_Vest_DisplayFrom_DisplayTo]
CHECK ([DisplayFrom] <= [DisplayTo])
GO
```

*e) Generated SQL script.*

*Figure 7. Consistent implementation of business rules (Vertical consistency).*

- **Productivity:** The productivity gains of SFs are twofold: The templates can be used to produce additional code without any further effort. Code base can be fully rebuilt to cover the changing solution requirements. Time to production is significantly reused, which can obviously improve customer satisfaction.

- **Abstraction:** SFs can store the system specification in a platform or language neutral form. Then this specification can be reused to rebuild the code base in a different language or with different design patterns [10]. System specification can contain business rules and can be written in a form which can be understood and reviewed by the business stakeholders.

- **Maintenance:** Code base is significantly reduced. Thus costs and complexity of maintenance are significantly reduced. The produced solution is more adaptable to the changing functional, non-functional and platform requirements. SF is also used to maintain and enhance the application.

- **Increased developers' satisfaction**: Developers are relieved from the repetitive type of coding since it is covered by the software factories. They can concentrate on those types of tasks which are more challenging, and require more creativity, and thus increase their job satisfaction.

- **Performance improvement**: Code produced by SF has significant performance improvement because it moves certain calculations from run-time to design and code production time, e.g. some logic and components can be implemented in the templates.

- **Support for aspect oriented programming**: XSLT templates can contain aspect advices and join cuts.

- **Improved testing and reliability**: Due to the consistency of implementation of business rules, if there is a problem with the artefact templates it will be discovered in the firstly tested entity. This reduces the amount of testing done on the produced application, and increases the reliability of the software application.

- **Support for rapid prototyping**: SF can be used to rapidly build an application prototype. Prototypes provide the customer with an early preview of the application, and with an opportunity to give an early feedback to the developers. This reduces the risk of misunderstandings between the developers and the customer.

- **Flatter class hierarchy**: Without SF, common functionality is implemented in parent classes from which more specific classes inherit. With SF, common functionality between classes is replicated by the code generator in all the classes where it is needed. Thus inheritance is more rarely used, and SF produces class hierarchy with lower number of levels. Such class hierarchies are easier to understand and modify.

We have also observed the following drawbacks of SFs:

- **Code Bloat**: There is a danger of code bloat. SF-produced code tends to be significantly larger than the hard-crafted code. However, one should note that teh SF-produced code should not be considered as source code for the purpose of time-and-cost estimation, software configuration management etc.

- **XSLT knowledge** is required. XSLT is not mainstream technology yet, and XSLT experienced developers are still hard to find.

- **Dual knowledge**: Developing artefact templates is a very complex task. Only very experienced programmers should be challenged with this task. Developers of templates need to be experienced and knowledgeable in both XSLT and the artefacts' programming language. For example, developing templates for SQL scripts requires knowledge of both XSLT and SQL. Debugging templates is lifts the debugging task to complexity levels not previously experienced by most of the programmers. Such programmers are difficult to find.

## 5. STATISTICS

Following table gives the percentage of SF generated and handcrafted code in the final fee calculation application for financial institutions.

| Handcrafted code | Automated code |
|---|---|
| 686KB (7%) | 9440KB (93%) |
| 148 Files (14%) | 915 Files (86%) |

*Table 1. SF generated vs. handcrafted code.*

The code that was produced by the SF is about 93% of the total code produced. The remaining 7% were hand coded because the corresponding business rules occurred less than 3 times. Similar percentages were observed also for the other web applications developed by the SF. Automated generation for 93% of the source code does not mean that the development time is reduced by 93%. Writing the templates is a complex programming task because these templates essentially contain the logic to write the code. Still, the improvements in productivity, consistency, quality, configurability, and functional extensibility are very significant. Depending on the requirements' complexity and their similarity with the already produced templates, application implementation can be cut by 50-90%. Similarly, the time to implement changes in user requirements can be cut by 50 to 90%. Similarly, the time to the first deliverables can be cut by half.

## 6. CONCLUSION

In this paper we elaborate the SF approach for rapid development of web applications. Although writing the artefact templates is a complex programming task because they essentially contain the logic to write the code, still the improvements in productivity, consistency, quality, configurability, and functional extensibility are very significant.

Our future work will concentrate around building an extension to the RoboCod Add-On which will allow for visual representation and manipulation of the DSL-based application model. Though this is only a non-functional improvement of our Software Factory, still it will make it more accessible even to less experienced programmers. Second, we will aim towards implementing support for more business rules thus making our SF more applicable to various business domains.